# Bi Substitution Effects on Superconductivity of Valence-Skip superconductor AgSnSe$_2$

Yudai Hijikata[1], Atsuhiro Nishida[1], Kohei Nagasaka[1], Osuke Miura[1], Akira Miura[2], Chikako Moriyoshi[3], Yoshihiro Kuroiwa[3], and Yoshikazu Mizuguchi[1]*

[1] *Department of Electrical and Electronic Engineering, Tokyo Metropolitan University, 1-1, Minami-osawa, Hachioji, Tokyo 193-0397, Japan*
[2] *Faculty of Engineering, Hokkaido University, Kita-13, Nishi-8, Kita-ku, Sapporo, Hokkaido 060-8628, Japan*
[3.] *Department of Physical Science, Hiroshima University, 1-3-1 Kagamiyama, Higashihiroshima, Hiroshima 739-8526, Japan.*

We have synthesized AgSn$_{1-x}$Bi$_x$Se$_2$ polycrystalline samples to investigate the effect of partial substitution of mixed-valence Sn by Bi$^{3+}$ to the superconductivity of the valence-skip superconductor AgSnSe$_2$. The Bi-substituted AgSn$_{1-x}$Bi$_x$Se$_2$ were obtained up to $x$ = 0.2, but an insulating phase Ag$_2$SnBi$_2$Se$_5$ with a NaCl-type structure showed up above $x$ = 0.3. The superconducting transition temperature increased from 4.5 K ($x$ = 0) to 5.0 K ($x$ = 0.1) by Bi substitution. The enhancement of superconductivity by the suppression of the valence-skip states of Sn suggests that the valence-skip states of Sn are not positively linked to the pairing mechanisms of superconductivity in the AgSnSe$_2$ system.

## 1. Introduction

Metal chalcogenides have been drawing much attention as high-transition-temperature (high-$T_c$), unconventional, and possible topological superconductors. Fe chalcogenides, FeSe and FeTe-based superconductors, have been actively studied as the structurally simplest Fe-based superconductor [1-5]. Notably, the $T_c$ can reach above 30 K by the external pressure effect [6-8] or metal/organic-ion intercalations [9,10]. As another layered superconductor family, BiS$_2$-based superconductors have been one of the recent topics in the field of new superconductors [11-13]. In addition, as possible topological superconductors [14,15], Bi$_2$Se$_3$-based and SnTe-based systems have drawn great attention [16-25]. As introduced here, metal-chalcogenide superconductors have been getting great interests as a playground in which novel superconductivity could emerge.

In this study, we focus on a metal-chalcogenide superconductor AgSnSe$_2$ [26,27]. The notable characteristic of this superconductor is the valence-skip states of Sn. Assuming the valences of Ag and Se as +1 and -2, respectively, the valence of Sn is formally expected as +3. The $Sn^{3+}$ state constitutes of the same amounts of $Sn^{2+}$ and $Sn^{4+}$ [26]. In such a valence-skip superconductor, the negative-$U$ mechanism can be expected to work, and a high $T_c$ can be expected [28], as emerged in $Ba_{1-x}K_xBiO_3$ (BKBO) [29]. However, the low $T_c$ of ~4.5 K with high carrier concentration of an order of $10^{22}$ $cm^{-3}$ was reported in $AgSnSe_2$ [27]. Hence, the negative-$U$ mechanism seems not to be working in the $AgSnSe_2$ superconductor.

Here, we have investigated the effect of partial substitution of mixed-valence Sn by $Bi^{3+}$ to the superconductivity of the valence-skip superconductor $AgSnSe_2$. Although the solubility limit of Bi for the Sn site in $AgSnSe_2$ was low (about 20%), we have observed the enhancement of the superconductivity by Bi substitution in $AgSn_{1-x}Bi_xSe_2$. The enhancement of superconductivity by the suppression of valence-skip states of Sn suggests that the valence-skip states of Sn are not positively linked to the pairing mechanisms of superconductivity in the $AgSnSe_2$ system.

## 2. Experimental details

Polycrystalline samples of $AgSn_{1-x}Bi_xSe_2$ were prepared by the solid-state-reaction method. The starting materials of Ag (99.9%) and Sn (99.99%) powders and Bi (99.999%) and Se (99.999%) grains were ground, mixed well, pressed into a pellet, and sealed into an evacuated quartz tube. After first sintering, the obtained pellet was ground and the same procedure was repeated to enhance the sample homogeneity. First, the synthesis temperature was optimized as 400 ºC (20 h) by synthesizing samples of $x = 0$ with several conditions. For all the compositions, the synthesis temperature of 400 ºC was used as well.

The phase purity of the obtained samples was preliminary examined by conventional powder X-ray diffraction (XRD) with a Cu-Kα radiation. The crystal structure of the samples and impurity phases were determined by powder synchrotron XRD with energy of 25 keV ($\lambda$ = 0.49657 Å) at the beamline BL02B2 of SPring-8 under a proposal No. 2016B1078. The synchrotron XRD experiments were performed at room temperature with a sample rotator system, and the diffraction data were collected using a high-resolution one-dimensional semiconductor detector MYTHEN with a step of $2\theta = 0.006º$. The crystal structure parameters were refined using the Rietveld method with RIETAN-FP [30]. The actual compositions of the obtained samples were analyzed using energy dispersive X-ray spectroscopy (EDX) with

TM-3030 (Hitachi). The crystal structure image was depicted using VESTA [31]

The temperature dependence of magnetizatuin was measured using a superconducting quantum interference device (SQUID) magnetometer with a typical applied field of 10 Oe after zero-field cooling (ZFC) and field cooling (FC). The temperature dependence of electrical resistivity was measured by the typical four-terminal method. Au wires were attached on the polished surface of the samples by using Ag pastes.

3. Results and discussion

On the basis of EDX analyses, we have confirmed that the Bi was homogeneously substituted in $x$ = 0.1 and 0.2, and the obtained Bi contents were almost the same as those refined using the Rietveld analyses. For $x$ = 0.5, the actual composition was determined as nearly Ag:Sn:Bi:Se = 2:1:2:5. As a fact, the solubility limit of Bi for the Sn site is near $x$ = 0.2, and the $Ag_2SnBi_2Se_5$ phase formed for $x$ = 0.3‑0.5.

Figure 1(a) shows the powder synchrotron XRD patterns for $x$ = 0–0.5. The crystal structure of the obtained samples was refined using the Rietveld method. For $x$ = 0, 0.1, and 0.2, XRD patterns were well refined using $AgSnSe_2$ structure model, a cubic NaCl-type model (*Fm*-3*m*: No. 225) with the atomic coordinates of Ag/Sn(0, 0, 0) and Se(0.5, 0.5, 0.5). Small impurity peaks of $SnSe_2$ (*P*-3*m*1: No. 164) and $Ag_8SnSe_6$ (*Pmn*$2_1$: No. 31) were detected. The XRD patterns and Rietveld fitting results for $x$ = 0, 0.1, and 0.2 are shown in Supplemental Materials (Fig. S1 and S2) [32]. The refined structure parameters and phase populations are listed in Supplemental Materials (Tables S1 and S2). For $x$ = 0 and 0.1, the reliable parameters $R_{wp}$ were 6.2% and 7.1%, indicating that the assumed structural model is appropriate. However, $R_{wp}$ for $x$ = 0.2 was 11.8%, which would be related to asymmetric peaks possibly attributed to another cubic phase with slightly different lattice parameters. Displacement parameter *B* for both sites clearly increased with increasing Bi amount (See Table S1.), suggesting local structural disorder increased by the Bi substitution of the Sn site.

As shown in Fig. 1(b), cleat phase separation is observed for $x$ = 0.3. The main phase is the NaCl structure with a larger lattice constant in which Ag, Sn and Bi randomly occupy cation sites: the atomic coordinates are Ag/Sn/Bi(0, 0, 0) and Se(0.5, 0.5, 0.5). Here after, we name $Ag_2SnBi_2Se_5$ phase. The other phase with a smaller lattice constant could not be refined. For $x$ = 0.4 and 0.5, the $Ag_2SnBi_2Se_5$ phase is stabilized. The XRD pattern for $x$ = 0.5 was refined using the $Ag_2SnBi_2Se_5$ model with a lattice constant of 5.81616(9) Å (with impurity phases of SnSe$_2$ (8%) and Ag$_8$SnSe$_6$ (3%)) as displayed in Supplemental Materials (Fig. S3). The obtained $R_{wp}$ was 11.0%. To our knowledge, the Ag$_2$SnBi$_2$Se$_5$ phase has not been reported so far. Assuming the valence states of Ag$^+$, Bi$^{3+}$, and Se$^{2-}$, the valence state of Sn is Sn$^{2+}$, which is consistent with the insulating transport properties observed for $x \geq 0.3$.

Figure 1(c) shows the Bi concentration dependence of lattice constant of $a$ for AgSn$_{1-x}$Bi$_x$Se$_2$ phase and the Ag$_2$SnBi$_2$Se$_5$ phase. The lattice constant of the Ag$_2$SnBi$_2$Se$_5$ phase is clearly larger than that of the AgSn$_{1-x}$Bi$_x$Se$_2$ phase.

Figure 2 shows the temperature dependences of magnetization (ZFC and FC) for $x = 0$, 0.1, 0.2, and 0.3. For $x = 0$–0.2, large shielding signals were observed, which indicates that observed superconducting states are bulk in nature. In contrast, the shielding signal of $x = 0.3$ was very small (almost zero shielding fraction). Therefore, the bulk superconductivity disappears for $x \geq 0.3$. These results are consistent with the formation of the Ag$_2$SnBi$_2$Se$_5$ phase for $x \geq 0.3$.

We defined $T_c^{onset}$ as the temperature of the onset in the magnetization curve and $T_{irr}$ (irreversible temperature) as the temperature where the ZFC and FC curves split. $T_c^{onset}$ was 4.7, 5.2, 5.1, and 4.0 K for $x = 0$, 0.1, 0.2, and 0.3, respectively. $T_{irr}$ was 4.4, 4.8, 4.9, and 3.3 K for $x = 0$, 0.1, 0.2, and 0.3, respectively. Basically, $T_c$ increases by Bi substitution for $x = 0$–0.2. In addition, for $x = 0.1$, the sharpness of the drop in the ZFC curve is the sharpest. Namely, the homogeneity of the emerging superconducting states is also enhanced by Bi substitution.

Figure 3(a) shows the temperature dependences of electrical resistivity ($\rho$) for $x = 0$, 0.1, and 0.2. Metallic behaviors were observed for $x = 0$ and 0.1. For $x = 0.1$, resistivity decreases as compared to $x = 0$. However, for $x = 0.2$, the resistivity increases, and the temperature dependence of $\rho$ becomes semiconducting-like, which indicates that the metallic conductivity is suppressed by the 20% Bi substitution. In addition, as shown in Supplemental Materials (Fig. S4), $\rho$ of the $x = 0.3$ sample is obviously high and exhibits a semiconducting temperature dependence. These results could be related to the formation of the Ag$_2$SnBi$_2$Se$_5$ phase. Although the XRD peak corresponding to the Ag$_2$SnBi$_2$Se$_5$ phase was not observed for $x = 0.2$, asymmetric diffraction peaks and the increase in the displacement parameter $B$ obtained by the Rietveld refinements may suggest that the insulating Ag$_2$SnBi$_2$Se$_5$ locally forms in the $x = 0.2$ sample as well.

Figure 3(b) shows the temperature dependences of $\rho$ at low temperatures for $x = 0$, 0.1, and 0.2. Zero-resistivity temperature $T_c^{zero}$ was 4.5, 5.0, and 4.8 K for $x = 0$, 0.1, and 0.2, respectively. The increase in $T_c$ by Bi substitution observed in the resistivity measurements is consistent with that observed in the magnetization measurements. From the magnetization and resistivity measurements, we confirmed that Bi substitution enhances the superconducting properties of $AgSnSe_2$, unless the insulating $Ag_2SnBi_2Se_5$ phase did not form.

In Fig. 4, the $T_c^{onset}$ ($M$) and $T_{irr}$ ($M$) estimated from the magnetization ($M$) measurements and $T_c^{zero}$ ($\rho$) estimated from the resistivity measurements are plotted to establish a superconductivity phase diagram of $AgSn_{1-x}Bi_xSe_2$. $T_c$ increases by Bi substitution and becomes the highest at $x = 0.1$. $T_c$ of $x = 0.2$ is still higher than that of $x = 0$. Basically, the Bi substitution positively works to enhancing superconductivity in $AgSnSe_2$. On the basis of the results, we can propose that the valence-skip states of Sn is not preferable for the emergence of superconductivity in $AgSnSe_2$. Then, the suppression of the valence-skip states by the addition of $Bi^{3+}$ ions enhances superconducting states. This conclusion is basically consistent with the previous study on the Sn substitution for the Ag site in $Ag_{1-x}Sn_{1+x}Se_2$ [26]: $T_c$ increases with increasing Sn content, which may result in the suppression of the valence-skip states of Sn, as well as the case of Bi substitution.

The effect of Bi substitution can also be detected by focusing the onset temperature in the temperature dependence of resistivity. As displayed in Supplemental Materials (Fig. S5), $T_c^{onset}$ ($\rho$) was 5.7, 6.3, and 9.7 K for $x = 0$, 0.1, and 0.2, respectively. Although the bulk $T_c$ ($T_c^{zero}$) decreases at $x = 0.2$ due to the formation of the $Ag_2SnBi_2Se_5$ phase, local $T_c$ still increases at $x = 0.2$. On the basis of these facts, we can expect a higher $T_c$ in heavily Bi-doped $AgSnSe_2$ than that in $x = 0.1$, if we could suppress the formation of the $Ag_2SnBi_2Se_5$ phase. Having considered the successful synthesis of Ag-doped SnTe (with a NaCl-type structure) by high-pressure synthesis [25] and the large lattice constant of $Ag_2SnBi_2Se_5$ than that of the $AgSn_{1-x}Bi_xSe_2$ system, the $AgSn_{1-x}Bi_xSe_2$ phase with a larger Bi amount can be stabilized by high pressure synthesis.

## 4. Conclusion

We have synthesized $AgSn_{1-x}Bi_xSe_2$ polycrystalline samples to investigate the effect of partial substitution of mixed-valence Sn by $Bi^{3+}$ to the superconductivity of the valence-skip superconductor $AgSnSe_2$. The Bi-substituted $AgSn_{1-x}Bi_xSe_2$ were obtained up to $x = 0.2$, but the insulating phase of $Ag_2SnBi_2Se_5$ with a NaCl-type structure showed up above $x = 0.3$. The bulk $T_c$ increased from $T_c^{zero} = 4.5$ K ($x = 0$) to 5.0 K ($x = 0.1$) by Bi substitution. In addition, $T_c^{onset}$ in the resistivity measurements, which is related to the local $T_c$, increased from 5.7 K ($x$

= 0) to 9.7 K ($x$ = 0.2). On the basis of the enhancement of superconductivity by Bi substitution in AgSnSe$_2$, we propose that the valence-skip states of Sn are not preferable for the emergence of superconductivity in the AgSnSe$_2$ system.

**Acknowledgment**

The authors would like to thank Dr. K. Kobayashi of Okayama University for fruitful discussion. This work was partly supported by Grant-in-Aid for Scientific Research KAKENHI (15H05886 and 16H04493).

*E-mail: mizugu@tmu.ac.jp

32) (Supplemental material) [Rietveld fitting of XRD patterns, refined structural parameters and phase populations, and temperature dependences of electrical resistivity] is provided online.

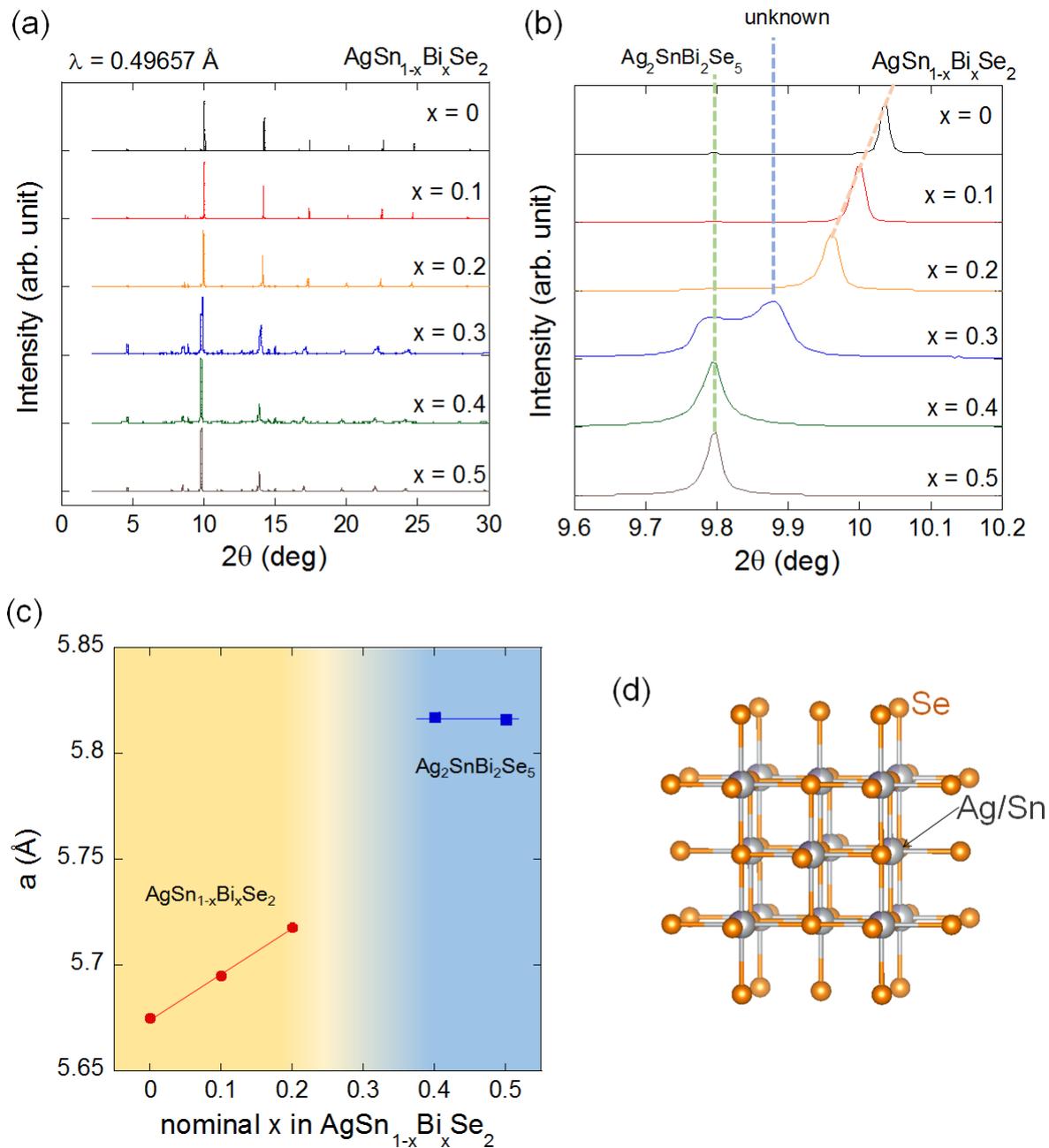

Fig. 1. (a) Synchrotron XRD patterns for $x = 0$–0.5. (b) Enlarged XRD patterns near (200) reflection. (c) Nominal $x$ dependence of lattice constant. (d) Schematic image of the crystal structure of $AgSnSe_2$. The crystal structure image was drawn using VESTA [31].

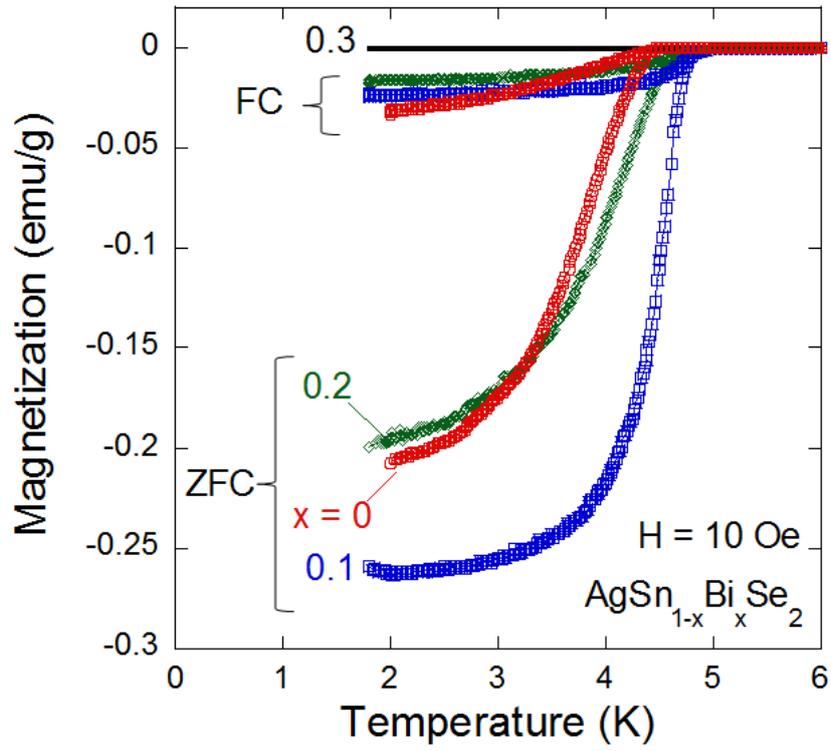

Fig. 2. Temperature dependences of magnetization (ZFC and FC) for $x$ = 0, 0.1, 0.2, and 0.3.

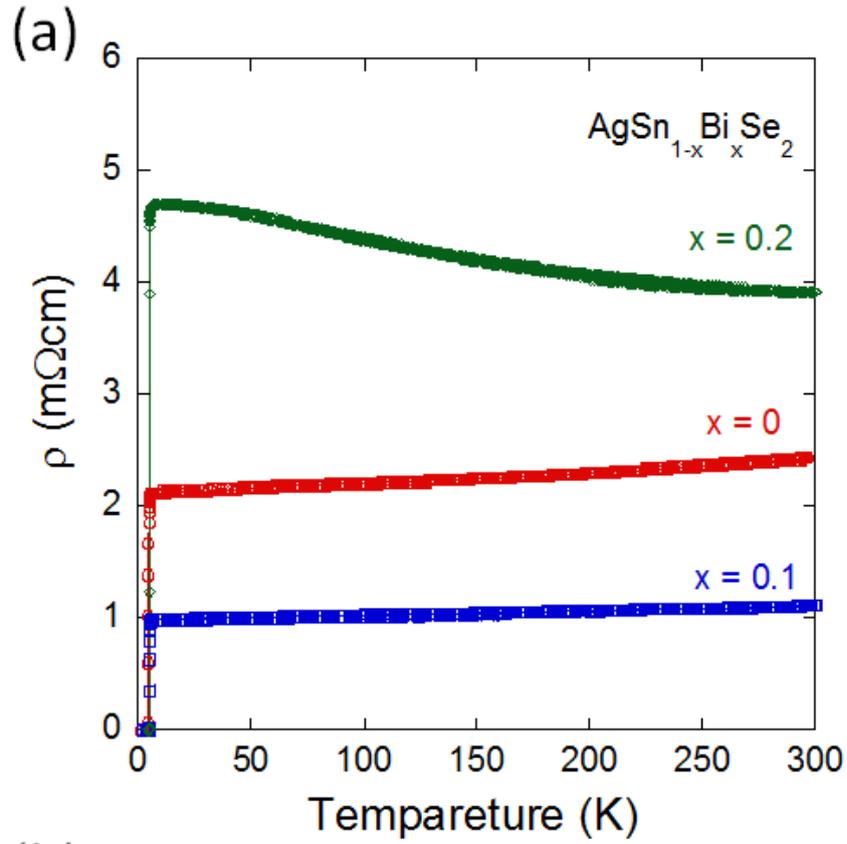

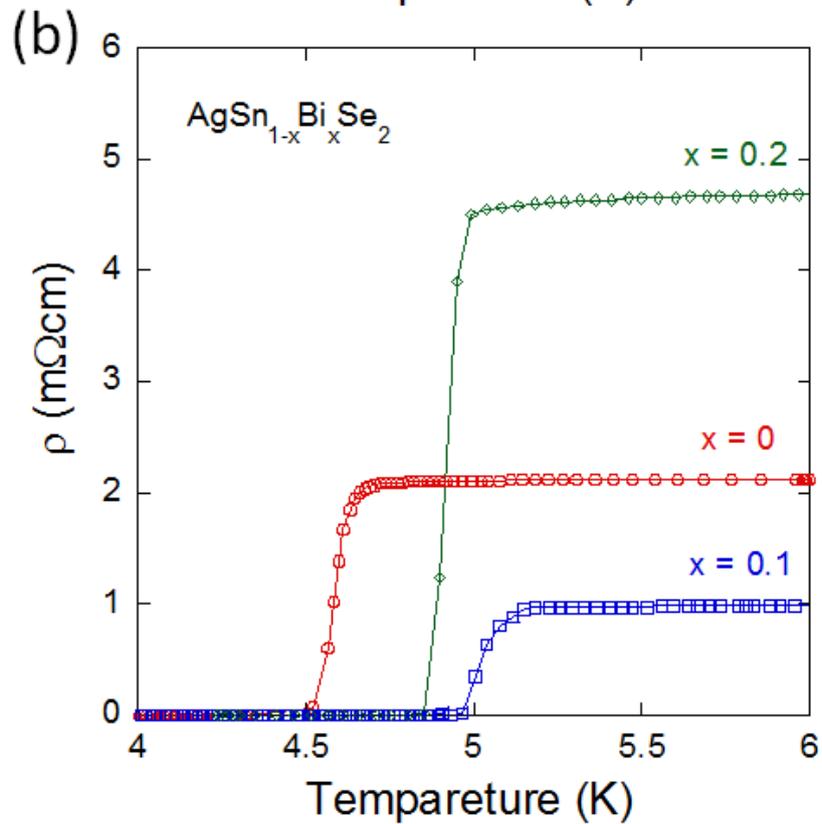

Fig. 3. (a) Temperature dependences of electrical resistivity ($\rho$) for $x$ = 0, 0.1, and 0.2. (b) Enlarged figures of the temperature dependences of $\rho$ at low temperatures for $x$ = 0, 0.1, and 0.2.

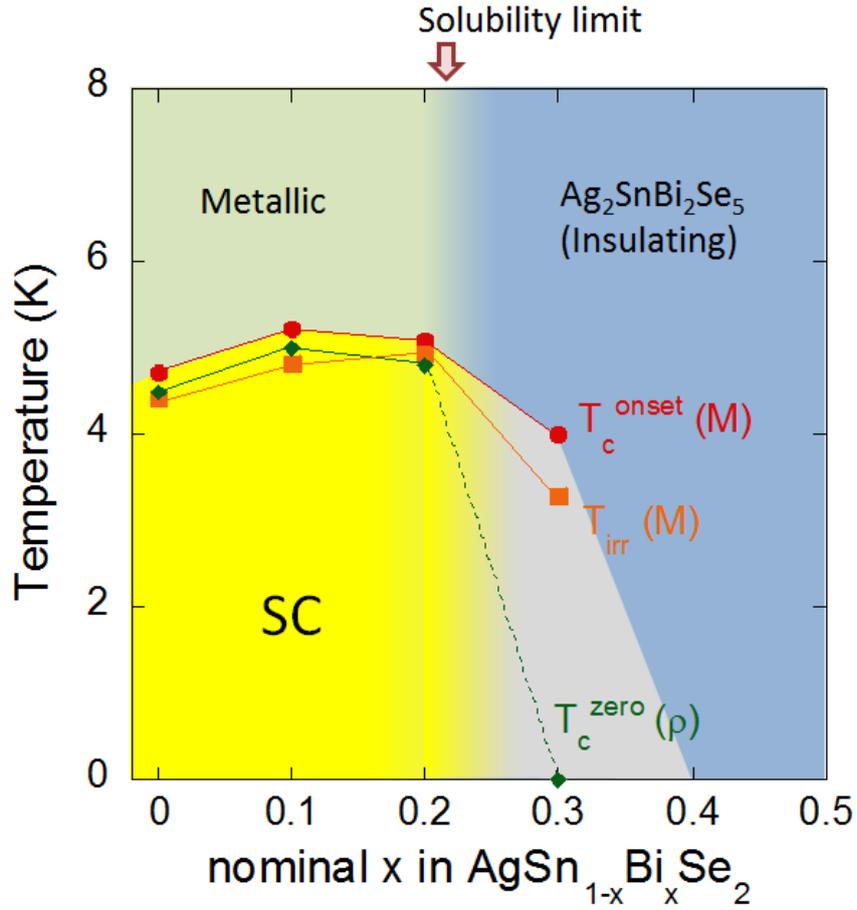

Fig. 4. Phase diagram of AgSn$_{1-x}$Bi$_x$Se$_2$. SC denotes a superconducting phase. $T_c^{onset}$ (M) and $T_{irr}$ (M) denote the onset temperature and the irreversible temperature in the magnetization measurements. $T_c^{zero}$ ($\rho$) denotes zero-resistivity temperature in the resistivity measurements.

# Supplemental Materials

Table S1. Crystal structure parameters for $x$ = 0, 0.1, and 0.2 refined using the Rietveld method. All parameters were refined using the NaCl-type model.

| $x$ (nominal) | 0 | 0.1 | 0.2 |
| --- | --- | --- | --- |
| $a$ (Å) | 5.67489(1) | 5.69526(1) | 5.71757(4) |
| $x$ (refined) | 0 | 0.12 | 0.2 |
| $B$ (Ag/Sn/Bi) (Å$^2$) | 1.262(8) | 1.35(1) | 1.55(3) |
| $B$ (Se) (Å$^2$) | 1.18(1) | 1.34(2) | 1.55(4) |
| $R_{wp}$ (%) | 6.2 | 7.1 | 11.8 |

Table S2. Phase populations for $x$ = 0, 0.1, and 0.2.

| $x$ (nominal) | 0 | 0.1 | 0.2 |
| --- | --- | --- | --- |
| AgSn$_{1-x}$Bi$_x$Se$_2$ (%) | 96 | 96 | 94 |
| SnSe$_2$ (%) | 3 | 3 | 4 |
| Ag$_8$SnSe$_6$ (%) | 1 | 1 | 2 |

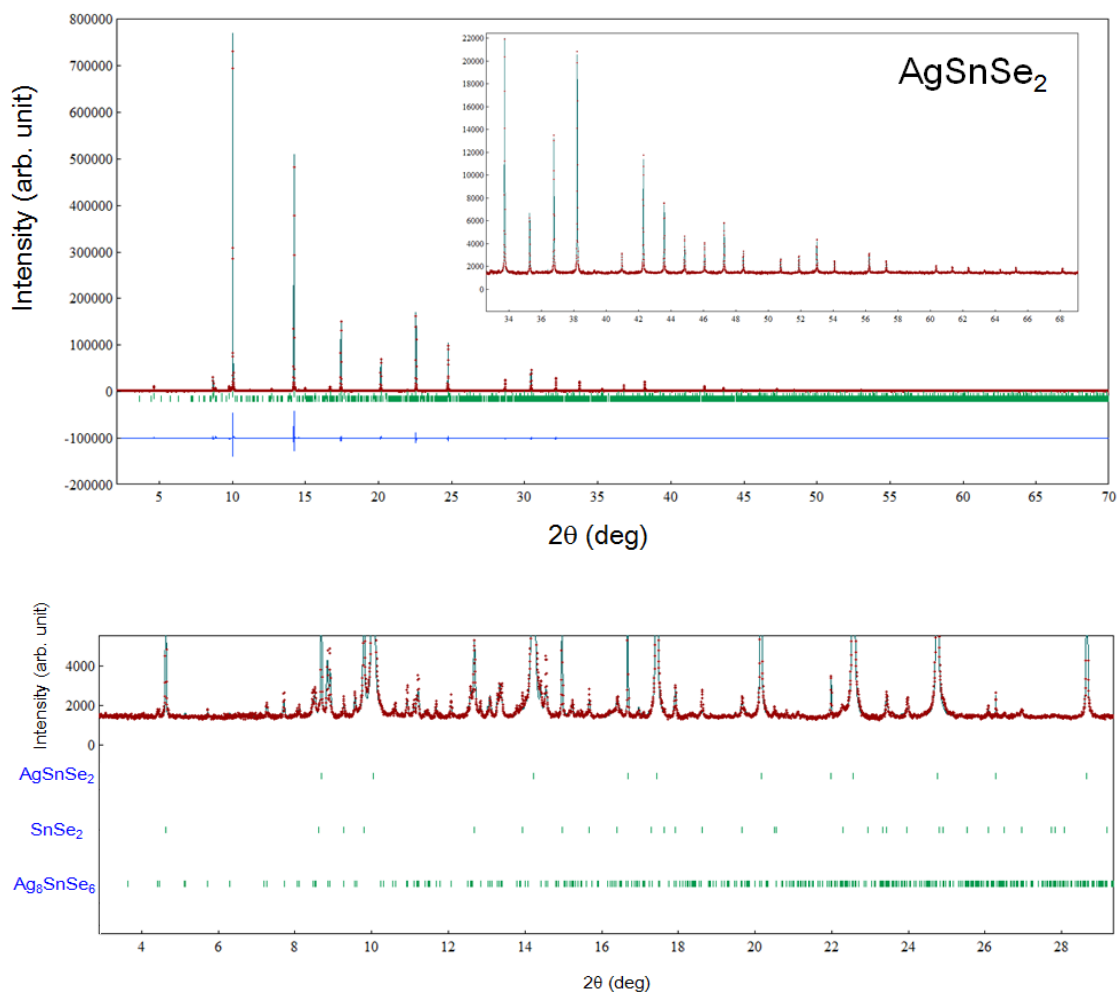

Fig. S1. Synchrotron XRD pattern and Rietveld fitting for $x = 0$ (AgSnSe$_2$). The higher-angle part is enlarged in the inset figure. An enlarged figure at low intensities is displayed in a lower panel.

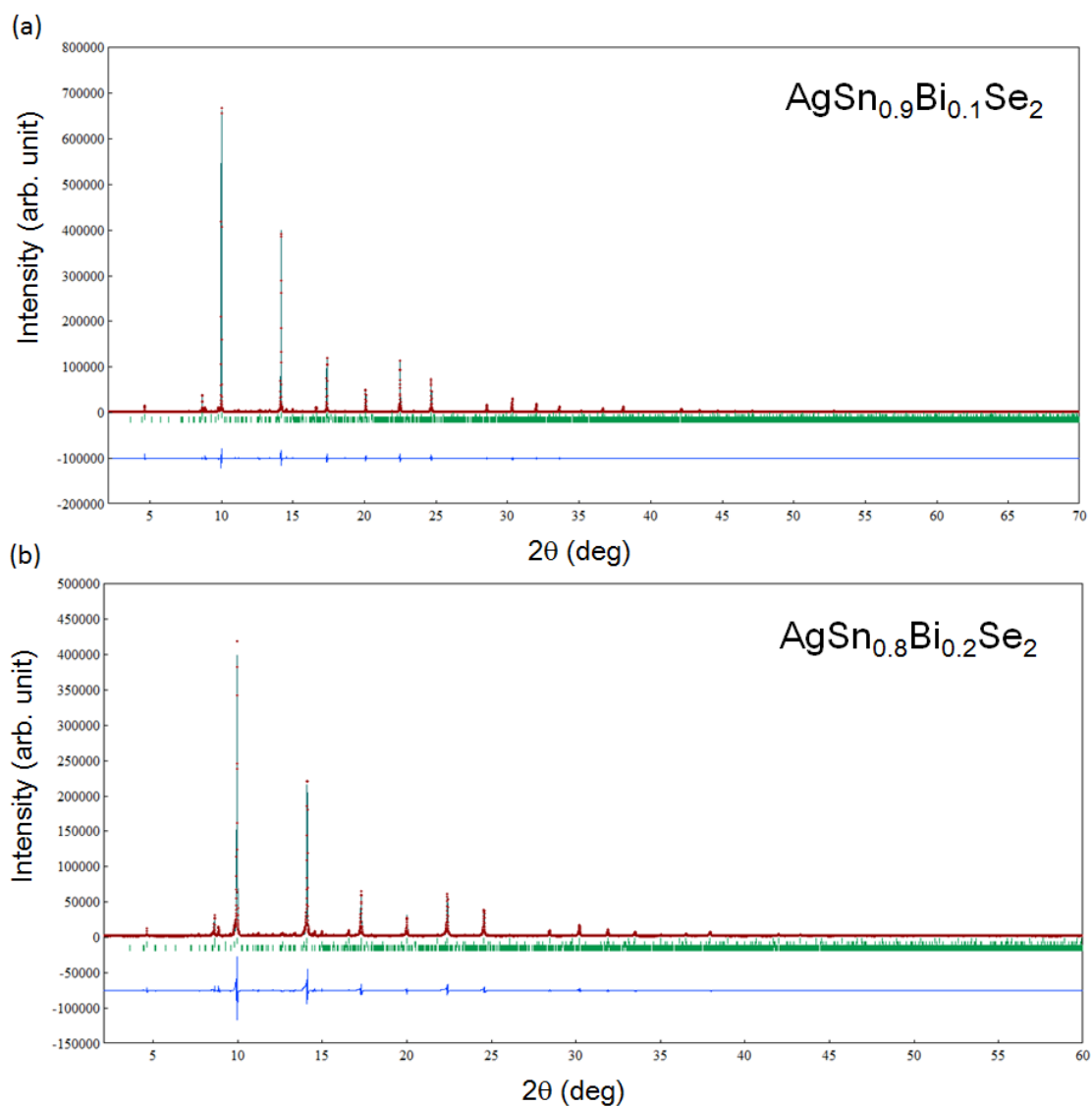

Fig. S2. (a,b) Synchrotron XRD pattern and Rietveld fitting for $x = 0.1$ ($AgSn_{0.9}Bi_{0.1}Se_2$) and 0.2 ($AgSn_{0.8}Bi_{0.2}Se_2$).

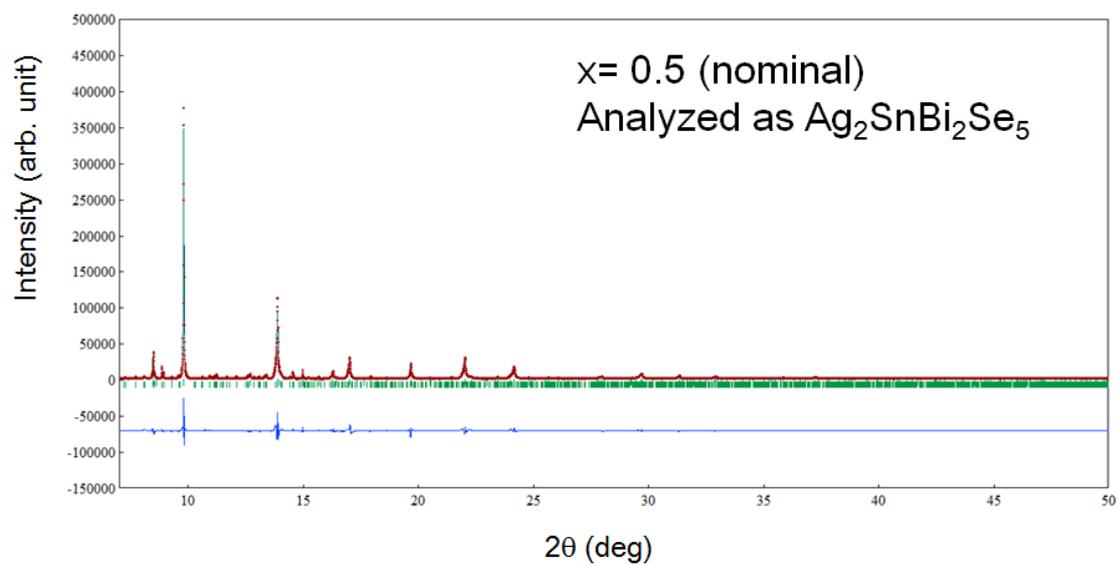

Fig. S3. Synchrotron XRD pattern and Rietveld fitting for $x = 0.5$. The Rietveld fitting was performed using a NaCl model with a composition of $Ag_2SnBi_2Se_5$.

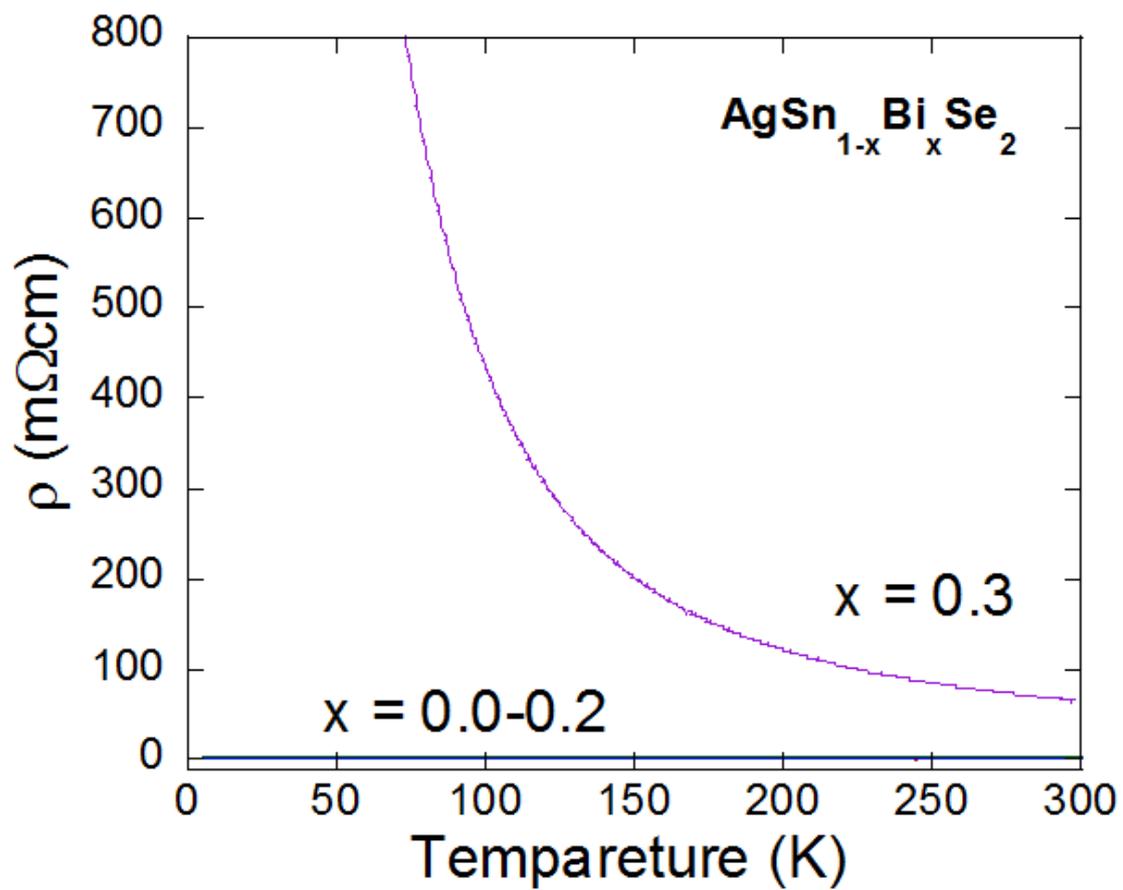

Fig. S4. Temperature dependence of electrical resistivity ($\rho$) for $x = 0.3$.

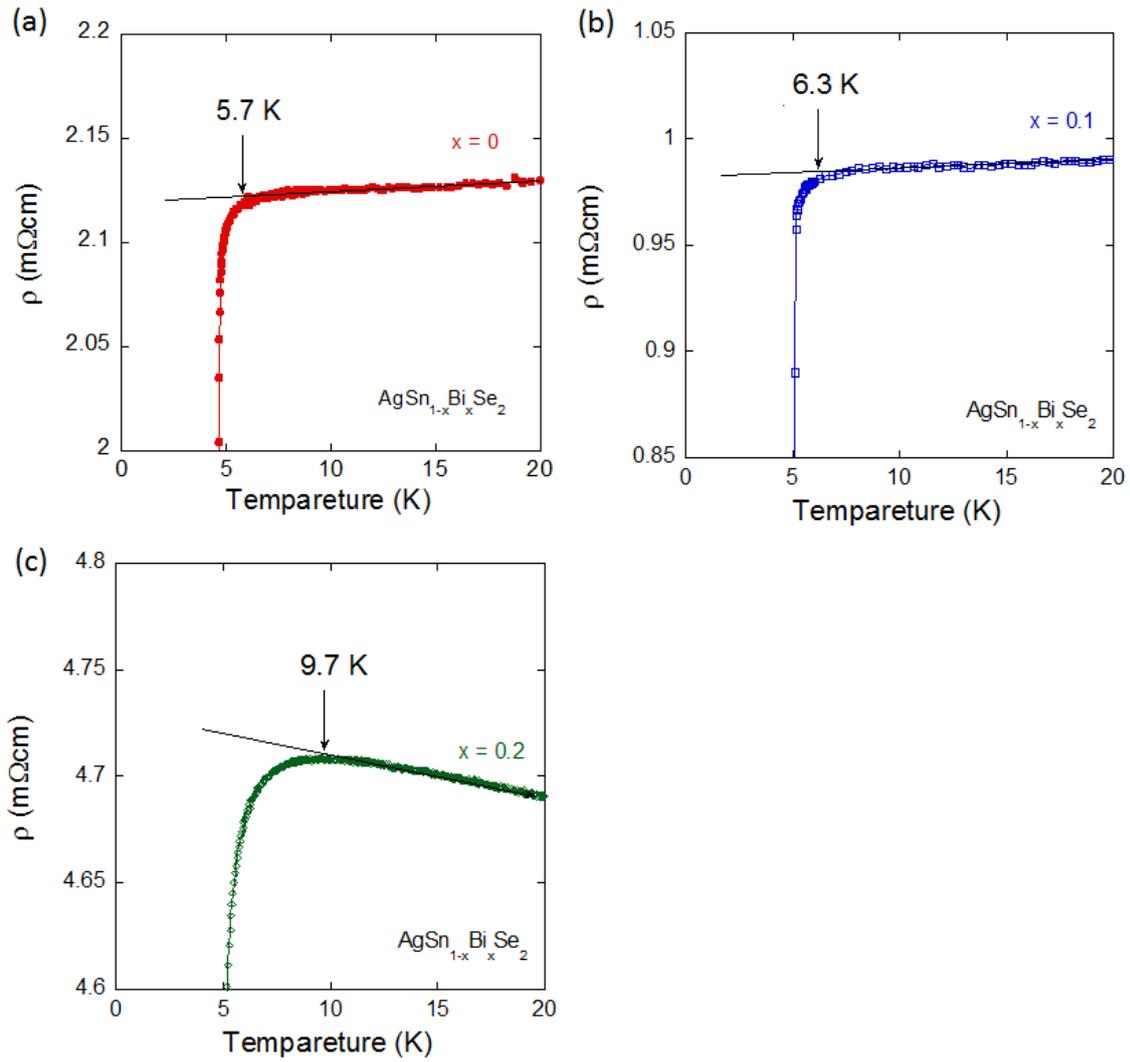

Fig. S5. Enlarged figures of the temperature dependence of electrical resistivity ($\rho$) for $x$ = 0, 0.1, and 0.2.